\documentclass[sigconf,natbib=true,anonymous=false]{acmart}
\usepackage{graphicx}
\usepackage{booktabs}
\usepackage{balance}
\usepackage{multirow}
\usepackage[normalem]{ulem}

\useunder{\uline}{\ul}{}

\AtBeginDocument{%
  \providecommand\BibTeX{{%
    \normalfont B\kern-0.5em{\scshape i\kern-0.25em b}\kern-0.8em\TeX}}}

\copyrightyear{2022}
\acmYear{2022}
\setcopyright{rightsretained}

\acmConference[SIGIR '22]{Proceedings of the 45th International ACM SIGIR Conference on Research and Development in Information Retrieval}{July 11--15, 2022}{Madrid, Spain.}
\acmBooktitle{Proceedings of the 45th Int'l ACM SIGIR Conference on Research and Development in Information Retrieval (SIGIR '22), July 11--15, 2022, Madrid, Spain}
\acmDOI{10.1145/3477495.3531737}
\acmISBN{978-1-4503-8732-3/22/07}

\usepackage{etoolbox}
\makeatletter
\patchcmd{\maketitle}{\@copyrightpermission}{
 \begin{minipage}{0.3\columnwidth}
  \href{http://creativecommons.org/licenses/by/4.0/}{\includegraphics[width=0.90\textwidth]{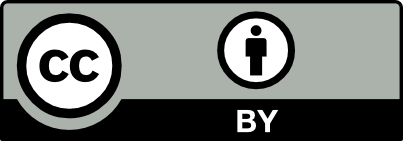}}
 \end{minipage}\hfill
 \begin{minipage}{0.7\columnwidth}
  \href{http://creativecommons.org/licenses/by/4.0/}{This work is licensed under a Creative Commons Attribution International 4.0 License.}
  \end{minipage}
  \vspace{5pt}
}{}{}
\makeatother

\begin{document}
\fancyhead{}


\title{ORCAS-I: Queries Annotated with Intent using Weak Supervision}


\author{Daria Alexander}
\affiliation{%
 \institution{Radboud University \& Spinque}
 \streetaddress{}
 \city{Utrecht}
 \state{}
 \country{The Netherlands}}
 \email{daria.alexander@ru.nl}

\author{Wojciech Kusa}
\affiliation{%
  \institution{TU Wien}
  \streetaddress{}
  \city{Vienna}
  \state{}
  \country{Austria}}
  \email{wojciech.kusa@tuwien.ac.at}

\author{Arjen P. de Vries}
\affiliation{%
  \institution{Radboud University}
  \streetaddress{}
  \city{Nijmegen}
  \state{}
  \country{The Netherlands}}
  \email{arjen@cs.ru.nl}



\begin{abstract}
User intent classification is an important task in information retrieval. In this work, we introduce a revised taxonomy of user intent. We take the widely used differentiation between navigational, transactional and informational queries as a starting point, and identify three different sub-classes for the informational queries: instrumental, factual and abstain. The resulting classification of user queries is more fine-grained, reaches a high level of consistency between annotators, and can serve as the basis for an effective automatic classification process. The newly introduced categories help distinguish between types of queries that a retrieval system could act upon, for example by prioritizing different types of results in the ranking.

We have used a weak supervision approach based on Snorkel to annotate the ORCAS dataset according to our new user intent taxonomy, utilising established heuristics and keywords to construct rules for the prediction of the intent category. We then present a series of experiments with a variety of machine learning models, using the labels from the weak supervision stage as training data, but find that the results produced by Snorkel are not outperformed by these competing approaches and can be considered state-of-the-art. The advantage of a rule-based approach like Snorkel's is its efficient deployment in an actual system, where intent classification would be executed for every query issued.

The resource released with this paper is the ORCAS-I dataset: a labelled version of the ORCAS click-based dataset of Web queries, which provides 18 million connections to 10 million distinct queries. We anticipate the usage of this resource in a scenario where the retrieval system would change its internal workings and search user interface to match the type of information request. For example, a navigational query could trigger just a short result list; and, for instrumental intent the system could rank tutorials and instructions higher than for other types of queries.


\end{abstract}




\maketitle

\section{Introduction}

When a user types a query into a search engine, there is usually a specific intent behind it: to download something, to purchase a product, to find a particular site or explore a topic. Understanding that intent can be very useful for providing relevant results to the searcher and increasing the value of the information obtained. Also, tailoring the content of the site to user intent helps increase the site's visibility rates.

While manual classification of user intent provides more accurate labels, manual annotation of large amounts of data can be very challenging. A weak supervision approach allows avoiding hand labelling of large datasets, which can save a lot of time and energy. In weak supervision, noisy labels are generated by using heuristics in the form of domain-specific rules or by using pattern matching. In this paper, we aim to understand how to automatically label click log data with user intent and annotate a large click dataset with those labels using weak supervision.

Commercial Web search engines refrain from disseminating detailed user search histories, as they may contain sensitive and personally identifiable information \cite{Adar07}. In the past, datasets such as AOL revealed personal information about the users, for example, their location and their names \cite{Barbaro06}. ORCAS \cite{Craswell20}, a new dataset released by Microsoft deals with that issue by not providing anything that could potentially help to identify the searcher. The absence of personal information and the large size of this dataset makes it very interesting for researchers, yet also makes impossible to analyze aspects like user behaviour during a search session.

Although many studies performed an automatic classification of user intent in search log data \cite{Lee05, Kang04, BaezaY06, Jansen08,Kathuria10,Lewandowski12}, there were fewer papers addressing this subject recently \cite{Figueroa15, Mohasseb19}. Also, to our knowledge there are no released large click datasets labelled with user intent. The datasets where user intent is annotated are mainly used for task-oriented dialogue systems. For example, MANtIS, a large-scale conversational search dataset containing more than 80,000 conversations across 14 domains that are englobing complex information needs \cite{Penha19}. Another dataset \cite{Larson19} was collected via crowd-sourcing and consists of 23,700 utterances covering 150 different intents. The Schema-Guided Dialogue dataset \cite{Rastogi20} has over 16,000 dialogues in the training set belonging to 16 domains. The intents in those datasets often differ from the intents for search log queries and are specific to interactions with conversational agents, such as ``transfer'', ``make payment'' and ``to do list update'' \cite{Larson19,Rastogi20}.

To fill this gap, we suggest using recent labelling techniques such as weak supervision combined with methods previously employed for intent classification of search log queries, such as a rule-based approach. We propose a user intent taxonomy based on a taxonomy established by Broder \cite{Broder02} that divides intents into three levels: informational, navigational and transactional. We perform the classification on two levels: 1) three categories of Broder's taxonomy, 2) three subcategories in the informational class: factual, instrumental and abstain. We base our automatic classification on Jansen et al.'s \cite{Jansen08} user intent characteristics, upon which we improve to further increase the quality of the taxonomy. Then we perform the labelling of the ORCAS dataset, which has 18 million connections to 10 million distinct queries. For the labelling, we use weak supervision with Snorkel \cite{Ratner2017}. After that, we train five different machine learning models on the 2 million items subset of the ORCAS dataset.

Our main findings are as follows:
\begin{itemize}
\item Our automatic labelling provides better results than those reported in the original study;
\item classifying the queries on the three top level categories provides better scores than classifying them on the full taxonomy;
\item benchmark models do not significantly outperform the Snorkel classifier;
\item the lack of performance of the models can be explained by 1) the specificities of weak supervision, 2) the lack of external knowledge such as ontologies, 3) the length of the queries and the absence of grammatical structure in them.
\end{itemize}

This work makes the following contributions:
\begin{itemize}
\item We improve the existing characteristics for automatic intent classification of search log queries;
\item we suggest subcategories that allow to have a more fine-grained automatic classification of user intent;
\item we share a publically available annotation for the widely used ORCAS dataset.

\end{itemize}

We release all three annotated versions of the ORCAS-I dataset \cite{kusa_alexander_devries_2022}.
Moreover, for reproducibility and transparency, we make our data labelling and classification scripts publicly available on GitHub\footnote{\url{https://github.com/ProjectDoSSIER/intents_labelling}}.

\section{Related work}

The related work relevant to this paper is linked to intent labelling, automatic classification of user intent and weak supervision.

\subsection{Intent labelling}

When users type queries in search engines, they often have a specific intent in mind. Broder \cite{Broder02} divides queries into three categories according to their intent: navigational, transactional and informational. An informational intent refers to acquiring some information from a website, a navigational intent consists of searching for a particular website, a transactional intent refers to obtaining some services from a website (e.g. downloading the game). In Broder's study, queries from AltaVista query log were classified manually and information about clicked URLs was not used. This taxonomy was expanded in \cite{RoseLevinson04} with sub-classes for informational, navigational and transactional categories. Contrary to Broder, clicked URLs were used for intent classification, but did not show significant improvement compared to labelling of queries only. The following studies used the complete Broder's taxonomy \cite{Jansen08,Kathuria10} or some of its categories \cite{Lee05,Kang04,Lewandowski11,Lewandowski12,Gul20}. Some studies added other categories, such as {\itshape browsing} \cite{Kellar07} or {\itshape learn} and {\itshape play} \cite{Russel09}.

\subsection{Automatic classification of user intent}
 
Early studies that performed automatic classification of user intent were usually limited to only two of Broder's categories: either informational and navigational \cite{Lee05,Kang04}, or informational and transactional \cite{BaezaY06}. They adopted different techniques such as computing the scores of distribution of query terms \cite{Kang04}, classification of queries into topics \cite{BaezaY06} as well as tracking past user-click behavior and anchor-link distribution \cite{Lee05}.

In order to automatically classify search intent, researchers used click features. They found that if the intent of a query was navigational, then users mostly clicked on a single website. On the other hand, if the intent was informational, users clicked on many results related to the query \cite{Lee05, Lewandowski12}. URL features, which take into account the text of the URLs were considered important for navigational category along with click features \cite{Lu06}. Also, using the text of the clicked URLs improved the results for the navigational category but not for the informational category \cite{Kang04}. 

Jansen et al. \cite{Jansen08} established a rule-based approach and defined query characteristics for automatic classification of informational, navigational and transactional intents. They were linked to query length, specific words and combinations of words encountered in queries, and to the information about the search session (e.g. whether it was the first query submitted). Assigning labels according to the established characteristics was done as a first step before using machine learning approaches, such as performing k-means clustering \cite{Kathuria10}. In order to add some additional features to Jansen et al.'s characteristics, natural language processing techniques such as POS-tagging \cite{Mohasseb19,Figueroa15}, named entity recognition and dependency parsing \cite{Figueroa15} were used, however, the classification was done on much smaller datasets than in the original study.

\subsection{Weak supervision}

One of the most common problems with successful training of machine learning models is the lack of datasets with good quality annotations. Manual collection of annotations is a costly, tedious and time-consuming process. Academic research institutions often do not have enough funding to gather large-scale annotations, limiting their capabilities of creating significant corpora. This became even more visible with the growth of large pre-trained language models \cite{Devlin2018,Brown2020} that led to impressive gains on many natural language understanding benchmarks \cite{wang2019superglue}, requiring a large number of labelled examples to obtain state-of-the art performance on downstream tasks \cite{zheng2021walnut}. In order to mitigate the lack of labelled data, recent works tried using other approaches to produce annotated datasets, like the usage of regular expression patterns \cite{augenstein-etal-2016-stance} or class-indicative keywords \cite{karamanolakis-etal-2019-leveraging}.

Weak supervision is an approach in machine learning where noisy, limited, or imprecise sources are used instead of (or along with) gold labelled data. It became popularised with the introduction of the data programming paradigm \cite{Ratner2016}. This paradigm enables the quick and easy creation of labelling functions by which users express weak supervision strategies or domain heuristics. Various weak supervision approaches can be represented by labelling functions, such as distant supervision, heuristics or the results of crowd-sourcing annotations. Weak supervision has already been successfully applied in other problems in the area of natural language processing and information retrieval \cite{badene2019weak,fries2019weakly,dehghani2017neural}. In this paper, we focus on the usage of heuristics to create the labelling functions for intent classification.

Snorkel is a weak supervision system that enables users to train models using labelling functions without hand labelling any data \cite{Ratner2017}. It is an end-to-end system for creating labelling functions and training and evaluating the labelling model. It is designed to work with classification or extraction tasks. According to the recent benchmark study \cite{zheng2021walnut}, it still offers comparable performance to newer and more complex weak supervision systems.

\section{Taxonomy}

Researchers do not agree on the terms to use for search behaviour classification. Some researchers use the notion of {\itshape intent} for determining informational, navigational and transactional search behaviour \cite{Broder02,Kathuria10,Jansen08}. Others use the term {\itshape goal} instead of the term {\itshape intent} for the same taxonomy \cite{Russel09,RoseLevinson04}. There is also a group of researchers who use the term {\itshape task} for the taxonomy that contains all \cite{Sellen02} or some \cite{Kellar07} of Broder's categories.

A search task is defined as a task that users need to accomplish through effective gathering of information from one or several sources \cite{Li10, BystromHansen05}. A task can have a goal, which is a part of a task description. A search intent is defined as the affective, cognitive, or situational goal as expressed in an interaction with information systems. \cite{Jansen08}.

As a search intent is an expression of a goal in an interaction with information systems and we analyse the data that
reflects this interaction, we decided to adopt Broder's choice of terms and use the notion of {\itshape intent}. For our study, we use Broder's initial intent classes: {\itshape informational}, {\itshape navigational} and {\itshape transactional}. We refine the informational class with three subcategories: 1) {\itshape factual} \cite{Kellar07,Jiang14}, 2) {\itshape instrumental} (also called {\itshape advice} in \cite{RoseLevinson04} or {\itshape learn} in \cite{Russel09}) and 3) {\itshape abstain}. 

\begin{figure}[h]
  \centering
  \includegraphics[width=0.75\linewidth]{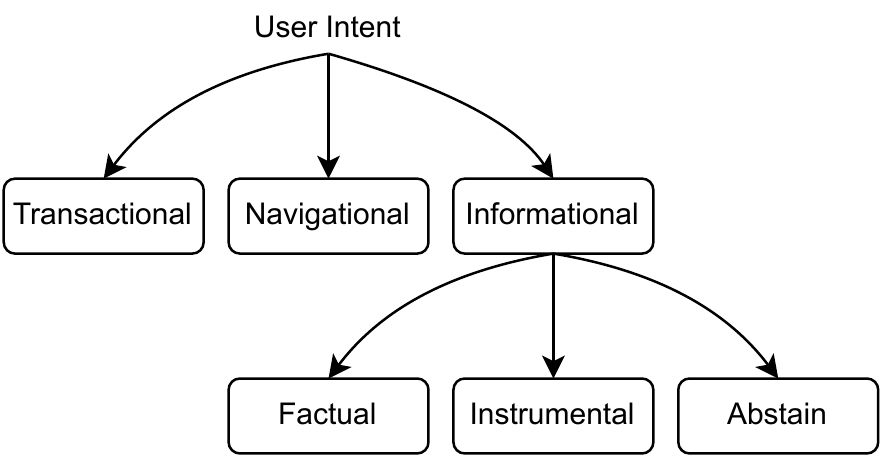}
  \caption{User intent taxonomy used in this study.} \label{fig:taxonomy}
  \Description{Graphical representation of our user intent taxonomy.}
\end{figure}

The {\itshape Factual} and {\itshape instrumental} subcategories were chosen because it was possible to identify characteristics that would allow their automatic classification and potentially many queries exist that would have those intents. We also considered other subcategories such as {\itshape descriptive} \cite{Kim06} and {\itshape locate} intents \cite{RoseLevinson04}, but found that they were to narrow for our goal of allowing classification. We provide the taxonomy (Figure \ref{fig:taxonomy}) as well as categories definitions. 

\begin{itemize}
\item \textbf{Navigational intent}: the immediate intent is to reach a particular website \cite{Broder02};
\item \textbf{Transactional intent}: locate a website with the goal to obtain some other product, which may require executing some Web service on that website \cite{Jansen08};
\item \textbf{Informational intent}: locate content concerning a particular topic in order to address an information need of the searcher \cite{Jansen08};
\subitem - \textbf{Factual intent}:  locate specific facts or pieces of information \cite{Kellar07};
\subitem - \textbf{Instrumental intent}: the aim is to find out what to do or how to do something \cite{Kim06};
\subitem - \textbf{Abstain}: everything inside the informational category that is not classified as factual or instrumental.
\end{itemize}

\section{Dataset}

To classify user intent of search queries according to Jansen's characteristics, previous studies used the Dogpile transaction log \cite{JansenSpink07} or the AOL web query collection \cite{Pass06}. Both of those datasets contained user IDs, which led to some privacy issues.

Concerned with those privacy problems, we decided to perform automatic annotation on a dataset that does not have user IDs. The ORCAS dataset appealed to us for 3 main reasons: 1) it does not contain information about users 2) it is a contemporary large dataset 3) it contains general queries such as one can find in any search engine. 

The ORCAS dataset is part of the MS MARCO datasets (Microsoft) and is intended for non-commercial research purposes. It contains 18.8 million clicked query-URL pairs and 10.4 million distinct queries. The dataset has the following information: {\itshape query  ID}, {\itshape query}, {\itshape document ID} and {\itshape clicked URL}. The documents that the URLs lead to come from TREC Deep Learning Track.

This dataset was aggregated based on a subsample of Bing's 26-month logs to January 2020. The creators of the dataset applied several filters to the log. Firstly, they only kept query-URL pairs where the URL is present in the 3.2 million document TREC Deep Learning corpus. Secondly, they applied a {\itshape k}-anonymity filter and only kept queries that are used by {\itshape k} different users. Finally, offensive queries such as hatred and pornography were removed.

For labelling, we use a 2-million sample of the ORCAS dataset. This is a sample that is chosen randomly from the whole dataset. We call this dataset \textbf{ORCAS-I-2M}. As the dataset is already pre-processed, we did not need to do any additional pre-processing. For example, the text of the queries is already lower cased. We decided to keep the punctuation in the queries because when the user is searching for a specific site, the dots before the domain names (such as ``.com'', ``.org'') can help to assign the right label to those queries. We also decided to keep multiple instances of the same query because they can potentially have a different label, depending on the contents of the URL clicked.

\section{Methodology}

In this section, we describe the process of creating the characteristics of user intent provided in our taxonomy. Those characteristics enable the automatic classification of the intent.

\subsection{Establishing characteristics for each label} \label{sec:characeristics}

The automatic assignment of labels to queries was based on the characteristics established by Jansen et al.\cite{Jansen08} for transactional, navigational and informational intents. However, we re-evaluated the characteristics for transactional and navigational categories and suggested new ones. Also, as we defined two subcategories (factual and instrumental) inside informational category, we decided to re-use some of Jansen et al.'s characteristics for this class and add new ones for each subcategory.

To determine user intent characteristics, queries drawn from different datasets (AOL web query collection \cite{Pass06}, TREC 2014 Session Track \cite{Carterette14}, MS MARCO Question Answering Dataset \cite{Tri16}) were analysed. For each characteristic we annotated small subsets of the datasets automatically and then for the next subsets we adjusted the characteristics of the classification; the characteristics that did not improve automatic labelling, for example, when they did not cover a significant part of the data, were discarded.

A big difference between our classification and Jansen's classification is that we do not only use queries but also URLs, as suggested by \cite{Lu06} and \cite{Kang04}. That helped us to refine the classification and include more features that could help to assign a label to a query.

We present the characteristics that we took from Jansen et al. as they were, those that we changed and those that we created ourselves. They are presented by category.
\\
\subsubsection{Transactional intent}~\\

We kept the majority of the characteristics from the transactional category. They are linked to various keywords that one would use when performing a transaction, such as ``download'', ``buy'', ``software''.

\noindent The characteristics we kept are:
\begin{itemize}
\item queries with ``download'' terms (e.g. ``download'', ``software'');
\item queries relating to image, audio, or video collections;
\item queries with ``audio'', ``images'', or ``video'' as the source;
\item queries with ``entertainment'' terms (pictures, games);
\item queries with ``interact'' terms (e.g. ``buy'', ``chat'');
\end{itemize}
The ones we did not use:
\begin{itemize}
\item queries with ``obtaining'' terms (e.g. lyrics, recipes);
\item queries containing terms related to movies, songs, lyrics, recipes, images, humor, and pornography; 
\item queries with compression file extensions (jpeg, zip).
\end{itemize}

As for the characteristics that we did not take, we empirically understood that 1) many queries that contain movies, songs, lyrics and recipes terms belong to the factual subcategory and 2) extensions such as ``jpeg'' and ``zip'' do not usually indicate transactional intent. For example, ``zip'' usually appears in phrase ``zip code'', which would belong to factual category. Also, many queries that contain the term ``jpeg'' would be classified under the instrumental category, such as ``converting to jpeg''.
\\
\subsubsection{Navigational intent}~\\

We kept two of Jansen et al. characteristics for navigational intent. For the queries containing domain names we used a list of top-level domains that we crawled from the Wikipedia\footnote{\url{https://en.wikipedia.org/wiki/List\_of\_Internet\_top-level\_domains}}.
\\

\noindent The characteristics we kept are:
\begin{itemize}
\item queries containing domains suffixes;
\item queries with ``Web'' as the source;
\end{itemize}
The ones we did not use:
\begin{itemize}
\item searcher viewing the first search engine results page;
\item queries length (i.e., number of terms in query) less than 3;
\end{itemize}
The ones that we refined: 
\begin{itemize}
\item queries containing company/business/organization/people names;
\item our version: queries for which the Levenshtein ratio between the query and the domain name is equal or greater than 0.55;
\end{itemize}

We did not take the characteristic of ``searcher viewing the first results page'', because, unlike Jansen et al., we do not have access to information about user sessions. We also decided that considering queries shorter than 3 words as navigational is counter-productive as 35\% of queries in the data have fewer than 3 words so it can potentially lead to many false positives. For example the queries ``allergic rhinitis'' and ``generation terms'', despite having only two words, do not belong to navigational category.

As for the ``queries containing company/organization/people names'', we found that navigational queries do not only contain the names of organizations and people. For example, the query ``army study guide'' leads to the site {\itshape https://www.armystudyguide.com} which is dedicated to US army study guide. Instead, we identified navigational intent by considering the similarities between the queries and the domain name parts of URLs. We used the Levenshtein similarity ratio which is calculated according to the following formula:

\begin{equation*}
  \frac{|a|+|b| - \text{Lev}(a,b)}{|a|+|b|}
\end{equation*}

Here, Lev$(a,b)$ is Levenshtein distance (the minimum number of edits that you need to do to change a one-word sequence into the other) and $|a|$ and $|b|$ are lengths of sequence {\itshape a} and sequence {\itshape b} respectively. A threshold on Levenshtein ratio was empirically established at 0.55, which means that if the query and the domain name were 55\% or more similar they were classified as navigational.
\\
\subsubsection{Informational intent}~\\

In his study, Jansen classified 81\% of queries as having informational intent. Follow-up research considered this category too broad \cite{Russel09}. It was one of the reasons that motivated us to introduce subcategories inside the informational category: factual, instrumental and abstain.

\paragraph{Factual intent}

Jansen et al.'s characteristics for informational intent such as having ``queries with natural language terms'' and ``queries length greater than 2'' are too broad to be useful. We did use ``queries that contain question words'' though. We suggest the following characteristics for factual intent:

\begin{itemize}
\item queries containing question words (e.g. ``what'', ``when'', ``where'', ``which'');
\item queries starting with question words (e.g. ``can'', ``does'');
\item queries containing words such as ``facts'', ``statistics'' and ``quantities'';
\item queries containing the terms linked to cost or price (e.g. ``average'', ``cost'', ``amount'', ``sum'', ``pay'');
\item queries containing words that can be replaced by numbers (e.g. ``phone'', ``code'', ``zip'');
\item queries containing words of definition (e.g. ``define'', ``definition'', ``meaning'');
\item the clicked URL leads to the sites that contain specific facts.
\end{itemize}

Usually, queries that contain question words require specific answers, ``what's the fastest animal in the world'', so the intent here is to find facts. After analysing search queries in different datasets, we realised that  queries that contain words associated with quantities, price and money have factual intent. Also, people searching for a term or concept definition usually look for specific information.

In order to get sites that provide specific facts, we took the 20 most frequent sites in the dataset and identified these sites among them (see Table \ref{tab:freq}). Usually those are encyclopedias or sites that contain some specific information (such as information about drugs, local weather etc.).

\begin{table}
  \caption{Most popular sites that provide specific facts in the 2M sample of ORCAS dataset.}
  \label{tab:freq}
  \begin{tabular}{ll}
    \toprule
    Name of the site&Count\\
    \midrule
    \ wikipedia.org&287,269\\
    \ webmd.com&23,110\\
    \ merriam-webster.com&19,881\\
    \ drugs.com&14,177\\
    \ dictionary.com&9,501\\
    \ mayoclinic.com&8,670\\
    \ reference.com&8,670\\
    \ britannica.com&7,894\\
    \ medicinenet.com&7,136\\
    \ accuweather.com&6,041\\
    \ weather.com&5,893\\
  \bottomrule
\end{tabular}
\end{table}

\paragraph{Instrumental intent}

No characteristics in Jansen et al. are relevant to instrumental intent, except ``queries that contain question words''. For the queries that are aimed at finding resources about what to do or how to do it, we established the following characteristics:

\begin{itemize}
\item queries containing question words (e.g. ``how to'', ``how do'', ``how does'');
\item queries that begin with infinitive form of a verb (e.g. ``build'', ``cook'');
\item queries that begin with the ``-ing'' form of the verb (e.g. ``making'', ``doing'');
\item the clicked URL leads to the sites that contain tutorials and instructions.
\end{itemize}

We figured out that the queries issued for finding advice or instructions often start with ``how to'' and ``how do''. Also, queries that start with a verb in the infinitive or using the ``-ing'' form usually have instrumental intent. For identifying the infinitives of the verbs we used a list of 850+ common English verbs\footnote{\url{https://github.com/ProjectDossier/intents_labelling/blob/main/data/helpers/verbs.txt}}. For queries that begin with the ``-ing'' form of a verb we chose {\itshape Spacy} \footnote{\url{https://spacy.io/}} to determine whether the part-of-speech of the first word is a verb and whether it has an ``-ing'' postfix. As well as for factual queries, we used the clicks to the sites among the 20 most popular sites, in this case to those that provide tutorials and advice.

\paragraph{Abstain category}

The queries that were not classified as transactional, navigational, factual or instrumental are assigned to abstain category. According to Jansen et al., the queries that do not meet criteria for navigational or transactional have informational intent. Thus, we decided to make abstain subcategory part of the informational category. However, we could not establish consistent automatic characteristics for this group of queries because we could not find any reliable patterns in them.

What are those abstain queries? We expect that many of these would belong to an \textit{exploratory} intent, when a user wants to learn or investigate something, but the goal of this search is amorphous \cite{Marchionini06,Jiang14}. Having user sessions usually helps to understand if the queries are exploratory. An exploratory search process is described as submitting a tentative query, then exploring the retrieved information, selectively seeking and passively obtaining cues about where the next steps lie \cite{White06}. As we do not have user sessions, establishing characteristics for those queries is infeasible for now.

\begin{table}
  \caption{Most popular sites that provide instructions and tutorials in the 2M sample of ORCAS dataset.}
  \label{tab:freq2}
  \begin{tabular}{ll}
    \toprule
    Name of the site&Count\\
    \midrule
    \ support.office.com&9,641\\
    \ support.apple.com&7,494\\
    \ wikihow.com&5,307\\
    \ support.google.com&4,348\\
  \bottomrule
\end{tabular}
\end{table}

\subsection{The test dataset}

In order to evaluate the performance of our weak supervision approach, we manually created a test set collection. We randomly selected 1000 queries from the original ORCAS dataset that were not in the ORCAS-I-2M dataset. The test set was annotated by two IR specialists using the open-source annotation tool {\itshape Doccano}\footnote{\url{https://github.com/doccano/doccano}}. In case of doubt about assigning a specific intent to a query, the result pages of clicked URLs were used as an additional hint for classification. For example, if the query intent was unclear and the result page was a tutorial, the query was classified as having instrumental intent. For inter-annotator agreement on the test set, the Cohen Kappa statistic was 0.82. Remaining disagreements between the two annotators were then resolved by discussion, leading to a final decision for every query.  We call this manually annotated dataset \textbf{ORCAS-I-gold}.

\subsection{Creating Snorkel labelling functions}

In machine learning terms, our intent taxonomy could be represented as a two-level, multi-class classification problem. Snorkel has originally only been implemented to handle annotations for single-level classification problems. As our taxonomy is hierarchical, we needed to define two layers of Snorkel labelling functions. 

We defined the first level of labelling functions to distinguish between navigational and transactional intents. All the queries that could not fit into one of these two categories were classified as informational intent in our taxonomy. Based on the characteristics defined in Section \ref{sec:characeristics}, we created four labelling functions for navigational queries and three functions for transactional queries.

On the second level, we defined labelling functions to cover factual and instrumental intents. Similar to the previous step, we designed nine factual functions and four instrumental labelling functions, using the characteristics from Section \ref{sec:characeristics}. All queries that were not assigned a label from the two layers of Snorkel got an abstain category.

We initially used {\itshape Spacy}'s \textit{en\_core\_web\_lg} language model to identify part of speech information and to detect named entities. After initial analysis, this proved to generate too many false negatives, especially for the detection of verbs. For example, the queries ``change display to two monitors'' and ``export itunes library'' were misclassified as abstain, because the verbs ``change'' and ``export'' were labelled as nouns. We suspect that these errors were primarily caused by the lack of a proper sentence structure, which prevented {\itshape Spacy} from correctly detecting the part of speech of the word. In the final version, we decided to use a list of the 850+ common verbs with which we obtained comparable coverage with fewer false positives. Eventually, we have only used {\itshape Spacy} for a labelling function where queries begin with the ``-ing'' form of the verb.

\subsection{Training Snorkel}

To obtain a final prediction score, we run independently two levels of Snorkel annotations. Based on our classification that all non-transactional, non-navigational queries are informational, for the second level prediction, we use all the queries which were assigned abstain from the first level. In order to conduct label aggregation, we experiment with both the LabelModel and MajorityLabelVoter methods implemented in Snorkel. LabelModel estimates rule weights using an unsupervised agreement-based objective. MajorityLabelVoter creates labels by aggregating the predictions from multiple weak rules via majority voting. We test their predictions on the test dataset using default hyperparameters. Results are presented in Table \ref{tab:snorkel-models-comparison}.

\begin{table}[ht]
\caption{Comparison of Snorkel labelling models results on ORCAS-I-gold.}
\label{tab:snorkel-models-comparison}
\begin{tabular}{@{}ccccc@{}}
\toprule
Model                      & Metric & Precision & Recall & F1-score \\ \midrule
\multirow{2}{*}{\begin{tabular}[c]{@{}c@{}}Majority Label\\ Voter\end{tabular}} & Macro avg       & {\ul .780}               & .763            & {\ul .771}              \\
                                    & Weighted avg    & {\ul .786}               & {\ul .783}            & {\ul .783}              \\
                                    \midrule
\multirow{2}{*}{Label Model}         & Macro avg       & .737               & {\ul .773}            & .750              \\
                                    & Weighted avg    & .779               & .770            & .772              \\ \bottomrule 
\end{tabular}
\end{table}

After analysis of results on the testset, MajorityLabelVoter achieved higher scores for all measures except for macro average recall. Therefore, we decided to use it to obtain the final labels for the ORCAS-I-2M dataset. MajorityLabelVoter has the additional benefit that it provides more explainable results, as for every query, the user can be presented with the raw aggregation of the single labelling functions.


\section{Benchmark models}

\begin{table*}
\centering
\caption{Accuracy of Snorkel classifier compared to other studies.}
\label{tab:other_studies}
\begin{tabular}{llcccc}
\toprule
             Study& Dataset & \# queries in the dataset & Features& Algorithm & Accuracy \\ \midrule
Jansen et al. \cite{Jansen08}&Dogpile transaction log&4,056,374& queries only &rules& 74\%\\
Ashkan et al. \cite{Ashkan09}&data from Microsoft adCenter&800,000& queries only&SVM& 86\% \\
Kathuria et al. \cite{Kathuria10}&Dogpile transaction log &4,056,374& queries only & k-means& 94\%\\
Figueroa \cite{Figueroa15}&data from the AOL query collection&4,811,638& queries and URLs & MaxEnt& 82.22\%\\ \midrule
\textbf{Our study}&ORCAS&18,823,602& queries and URLs &rules& 90.2\%\\ 
\bottomrule
\end{tabular}%
\end{table*}

We benchmark five different models by training them on the ORCAS-I-2M dataset and evaluating the results on ORCAS-I-gold. We split the ORCAS-I-2M dataset into train and validation sets with 80:20 ratio. Hyperparameters not mentioned below are given their default values.

\begin{itemize}
    \item \textbf{Logistic regression}: For logistic regression we use tf-idf for text representation and the sklearn \cite{scikit-learn} standard scaler for feature scaling.
    \item \textbf{SVM}: Support Vector Machine. As our dataset is large, we use linear support vector classification. As for logistic regression, we use tf-idf vector for text representation.
    \item \textbf{fastText}: We use the fastText \cite{Bojanowski2016} model with word embeddings trained from scratch on ORCAS-I-2M dataset. This is a text classification model that uses average of word embeddings to compute the representation of a document. We utilise the Python wrapper implementation\footnote{\url{https://pypi.org/project/fasttext/}}.
    \item \textbf{BERT}: We use pre-trained, 110M parameters BERT model \cite{Devlin2018} followed by a classification head. We use the \textit{bert-base-uncased} model implemented in the HuggingFace library \cite{wolf2019huggingface}. Batch size is set to 64, fine-tuning is conducted for 10 epochs.
    \item \textbf{xtremedistil}: We also evaluate \textit{xtremedistil-l6-h384-uncased} checkpoint from the  XtremeDistilTransformers model \cite{mukherjee2021xtremedistiltransformers}. Similar to BERT, we fine-tune it for 10 epochs with a batch size set to 64.
\end{itemize}

\section{Results}

In this section, we present the statistics of ORCAS-I annotated both manually and with Snorkel. We also show the results of training the benchmark models on ORCAS-I-2M when evaluated on ORCAS-I-gold.

\subsection{Snorkel classification}

\begin{table}[htbp]
\caption{Detailed Snorkel weak labelling results for the test set using only three top level categories on ORCAS-I-gold.}
\label{tab:snorkel-first-lev}
\begin{tabular}{@{}ccccc@{}}
\toprule
Category      & Precision & Recall & F1-score & Examples \\ \midrule
Navigational  &.776      &.731   &.753    &    171  \\
Transactional &.756      &.791   &.773    &     43 \\ 
Informational &.936      &.945   &.941    &    786  \\ \midrule
Macro Avg     &.823      &.822   &.822    &   1000   \\
Weighted avg  &.901      &.902   &.901    &   1000 \\ \midrule \midrule
&\multicolumn{2}{c}{Accuracy}                        &.902 & 1000     \\ \bottomrule
\end{tabular}
\end{table}

\begin{table}[htbp]
\caption{Detailed results for Snorkel weak labelling for full intent taxonomy on ORCAS-I-gold.}
\label{tab:snorkel-both-levels}
\begin{tabular}{@{}ccccc@{}}
\toprule
Category         & Precision & Recall & F1-score & Examples \\ \midrule
Navigational  &.800      &.725   &.761  & 171   \\
Transactional &.756      &.791   &.773 &  43  \\ 
Instrumental  &.774      &.695   &.732  &  59  \\
Factual       &.847      &.826   &.837  & 363   \\ 
Abstain       &.723      &.780   &.750  & 364  \\ \midrule
Macro avg     &.780      &.763   &.771 &  1000  \\
Weighted avg  &.786      &.783   &.783 &  1000  \\ \midrule \midrule
&\multicolumn{2}{c}{Accuracy}                         &.783 &  1000  \\ \bottomrule
\end{tabular}
\end{table}

\subsubsection{Top level categories} We run the Snorkel classifier on the ORCAS-I-gold testset. Table \ref{tab:snorkel-first-lev} shows that it attains an F1-score of 0.82 and an accuracy of 0.90.  The category with the best performance is informational, followed by transactional and navigational. 

Table \ref{tab:other_studies} shows how our approach outperforms the results of the original Jansen et al. paper and also those reported in other studies, except \cite{Kathuria10}, that reaches an accuracy of 0.94. However, the informational category is over-represented in the ORCAS-I-gold, as well as in ORCAS-I-2M dataset (see section 7.4 and Table \ref{tab:label-distribution}), which can potentially influence the quality of training of the models such as BERT.

\subsubsection{Full taxonomy}

Table \ref{tab:snorkel-both-levels} shows that in the full intent taxonomy, the factual subcategory has the best results. It is followed by navigational category, which gets slightly higher precision results compared to predictions from three top level categories. The transactional and abstain categories as well as the instrumental subcategory perform worse than the others. For the transactional and instrumental categories this result can be linked to the small number of queries of this type in ORCAS-I-gold.

The results show that using more categories lowers the overall F1-score and accuracy. We can conclude that having more classes and more rules can potentially diminish Snorkel performance. Nevertheless, having subcategories in the informational category would allow to provide more choice for distinguishing between different types of intent.

\subsection{Benchmark models}

\begin{table}[htbp]
\caption{Macro average scores comparison for all benchmark models trained on three top level categories. Underlined scores indicate the highest score within the different input features for each model, bold values indicate the highest score overall.}
\label{tab:benchmarks-orcas-gold-first-level}
\begin{tabular}{@{}clccc@{}}
\toprule
Model                                                                           & \multicolumn{1}{c}{Input features} & Precision   & Recall      & F1-score    \\ \midrule
\multirow{2}{*}{Snorkel}                                                                         & query                         & {\ul \textbf{.864}} & .700 & .742\\ 
                                                                                & query + URL                        & .822 &  {\ul .822}& {\ul .822} \\ \midrule \midrule
\multirow{2}{*}{\begin{tabular}[c]{@{}c@{}}Logistic \\ regression\end{tabular}} & query                              & .796       & .756       & .770       \\
                                                                                & query + URL                        & {\ul.815} & {\ul.758} & {\ul.784} \\ \midrule
\multirow{2}{*}{SVM}                                                            & query                              & {\ul.842} &.783       &.798       \\
                                                                                & query + URL                        &.824       & {\ul.817} & {\ul.817} \\ \midrule
\multirow{2}{*}{fastText}                                                       & query                              &.788       &.737       &.748       \\
                                                                                & query + URL                        & {\ul.820} & {\ul.801} & {\ul.806} \\ \midrule
\multirow{2}{*}{BERT}   & query                              &.821           &.785           & .796           \\
                                                                                & query + URL                        & {\ul.832}           & {\ul \textbf{.823}}           & {\ul \textbf{.826}}           \\ \midrule
\multirow{2}{*}{xtremedistil}   & query                              & {\ul.846}           &.771           &.790           \\
                                                                                & query + URL                        & .818           & {\ul.818}           & {\ul.817}           \\ \bottomrule
\end{tabular}
\end{table}

\begin{table}[htbp]
\caption{Macro average scores comparison for all benchmark models trained on all the categories. Underlined scores indicate the highest score within the different input features for each model, bold values indicate the highest score overall.}
\label{tab:benchmarks-orcas-gold}
\begin{tabular}{@{}clccc@{}}
\toprule
Model                                                                           & \multicolumn{1}{c}{Input features} & Precision            & Recall               & F1-score             \\ \midrule
\multirow{2}{*}{Snorkel}                                                                         & query                         & .771 & .648 & .667          \\ 
                                                                      & query + URL                        & {\ul.779}          & {\ul.764}          & {\ul.770}          \\ \midrule \midrule
\multirow{2}{*}{\begin{tabular}[c]{@{}c@{}}Logistic \\ regression\end{tabular}} & query                              & .701                & .611                & .643                \\
                                                                                & query + URL                        & .{\ul714}                & {\ul.689}                & {\ul.700}                \\\midrule
\multirow{2}{*}{SVM}                                                            & query                              & .735                & .689                & .703                \\
                                                                                & query + URL                        & {\ul.782}          & {\ul.759}          & {\ul.767}          \\ \midrule
\multirow{2}{*}{fastText}                                                       & query                              & .694                & .643                & .660                \\
                                                                                & query + URL                        & {\ul.768}          & {\ul .753}          & {\ul .758}          \\ \midrule
\multirow{2}{*}{BERT}   & query                              & .742                & .705                & .717                \\
                                                                                & query + URL                        & {\ul \textbf{.789}} & {\ul .764}               & {\ul \textbf{.774}} \\ \midrule
\multirow{2}{*}{xtremedistil}   & query                              & .725                & .691                & .696                \\
                                                                                & query + URL                        & {\ul.781} & {\ul \textbf{.765}}               & {\ul.772} \\ \bottomrule
\end{tabular}
\end{table}

To train the models on ORCAS-I-2M, we use two types of training data: just the query and query plus URL. URL features help improve classification effectiveness. Tables \ref{tab:benchmarks-orcas-gold-first-level} and \ref{tab:benchmarks-orcas-gold} show that when we eliminate URL features from Snorkel (we mute or change the labelling functions that are using URLs) especially recall is reduced. This is particularly noticeable for the navigational category, for which recall drops from 0.73 to 0.35. This confirms the findings of \cite{Kang04} that using the text of clicked URL improves the results for navigational category. 

We hypothesise that as we take URL features into account for the Snorkel classifier, models that train on queries and URLs will outperform the models that train on queries only. This hypothesis is confirmed for the full taxonomy, especially for fastText and xtremdistil. For the three top level categories, the difference in performance is not so large (except for fastText), which can be explained by fewer URL features for these categories.

Even if we only use the query, the recall for the models trained on the three top level categories is higher than the recall of Snorkel without URL functions. As for the full taxonomy, SVM, BERT and xtremdistil show improvements on recall for query-only when compared to Snorkel. It indicates that the models learn well from the labels assigned by the Snorkel query and URL functions, even if they are trained on queries only.

None of our benchmark models significantly outperforms our Snorkel baseline when trained on queries and URLs. This could have been an expected behaviour when comparing two models, one being the teacher and the other the student who learned only from this one teacher, without any external knowledge. We also hypothesise that transformer-based models cannot express their full power because the input sequences are, on average, very short and often do not have a proper grammatical structure.

While our experiment does not indicate how the machine learning models could strictly outperform the base, weak labelling, we see some potential directions for future work. One solution would be to use more than one annotated click log dataset, ideally using different annotation types (i.e. a either combination of weak and human annotations or two distinct weak supervision models). Another solution would be to use a model that could utilise an external knowledge base or ontology to understand the nuances between different categories, which often depend on the type of website that the user selected.

\subsection{Final intent classification}

After analysing results on ORCAS-I-2M for both Snorkel and other benchmark models we decided to use the Snorkel model to predict intent categories on the full ORCAS dataset. We call this dataset \textbf{ORCAS-I-18M}. As ORCAS-I-18M contains all items that were included in both ORCAS-I-gold and ORCAS-I-2M, it should be used with caution when training and evaluating machine learning models.

\subsection{Dataset statistics}

Overview statistics of all ORCAS-I datasets used in this study are presented in Table \ref{tab:general-stats}. ORCAS-I-2M covers more than half of the unique domains available in ORCAS-I-18M, and at the same time, more than 40\% of unique URLs. This means that even though ORCAS-I-2M constitutes only around 10\% of the ORCAS-I-18M elements; it is still a representative sample. The mean length of the query is comparable in all ORCAS-I datasets. We noticed that 246 duplicated query-URL pairs exist in the ORCAS-I-18M dataset and 7 in ORCAS-I-2M. Even though we do not preserve uniqueness in our training data, such a small amount of duplicates would not affect the training of machine learning models.

\begin{table}[htbp]
\caption{Statistics of ORCAS datasets used in this paper (``un.'' stands for ``unique'').}
\label{tab:general-stats}
\begin{tabular}{crrr}
\toprule
                                                                            & ORCAS-I-gold & ORCAS-I-2M  & ORCAS-I-18M \\ \midrule
dataset size                                                         & 1,000 & 2,000,000 & 18,823,602       \\
un. queries                                                              & 1,000 & 1,796,652 & 10,405,339       \\
un. URLs                                                                 & 995  & 618,679   & 1,422,029        \\
un. domains                                                              & 700    & 126,001   & 241,199        \\ \midrule
\begin{tabular}[c]{@{}c@{}}un. words \\ in query\end{tabular}                                                              & 2,005    & 334,724   & 1,380,160        \\ 
\begin{tabular}[c]{@{}c@{}}mean query \\ length (words)\end{tabular} & 3.21       & 3.25      & 3.25       \\ 
\bottomrule
\end{tabular}
\end{table}

We measure the label distribution for all three annotated ORCAS-I datasets and present the results in Table \ref{tab:label-distribution}. To test the quality of Snorkel, we compare the distribution on ORCAS-I-gold both for manual annotations and the output from Snorkel weak labelling. We notice, that the only underrepresented category from our weak supervision is the navigational intent, which contains 1.6\% less items than in manual labelling approach. These queries were mostly categorised as abstain by our weak supervision. The label distribution from Snorkel for all three datasets is comparable, so both gold and 2M samples chosen from the full ORCAS dataset are representative.

\begin{table}[htbp]
\caption{Label distribution for all three annotated ORCAS-I datasets.}
\label{tab:label-distribution}
\begin{tabular}{@{}lcccc@{}}
\toprule
\multicolumn{1}{c}{\multirow{2}{*}{\begin{tabular}[c]{@{}c@{}}Label \\distribution\end{tabular}}} & \multicolumn{2}{c}{ORCAS-I-gold}                                  & ORCAS-I-2M                        & ORCAS-I-18M                      \\ 
                                    & Manual                         & Snorkel                        & Snorkel                               & Snorkel                               \\ \midrule
Navigational                           & 17.10\%                        & 15.50\%                        & 14.48\%                        & 14.51\%                        \\ \midrule
Transactional                          & 4.30\%                         & 4.50\%                         & 4.17\%                       & 4.16\%                         \\  \midrule
Informational                          & 78.60\%                        & 80.00\%                        & 81.35\%                        & 81.33\%                        \\
\hspace{0.1em}{\color[HTML]{555555}- Instrumental}                           & {\color[HTML]{555555} 5.90\%}  & {\color[HTML]{555555} 5.30\%}  & {\color[HTML]{555555} 5.82\%}  & {\color[HTML]{555555} 5.81\%}  \\
\hspace{0.1em}{\color[HTML]{555555}- Factual}                                & {\color[HTML]{555555} 36.30\%} & {\color[HTML]{555555} 35.40\%} & {\color[HTML]{555555} 35.35\%} & {\color[HTML]{555555} 35.35\%} \\
\hspace{0.1em}{\color[HTML]{555555}- Abstain}                                & {\color[HTML]{555555} 36.40\%} & {\color[HTML]{555555} 39.30\%} & {\color[HTML]{555555} 40.18\%} & {\color[HTML]{555555} 40.17\%} \\ \bottomrule
\end{tabular}
\end{table}

\section{Conclusion}

In this paper we revise the taxonomy of user intent, using the widely used classification of queries into navigational, transactional and informational as a starting point. We identify three different sub-classes for the informational queries: instrumental, factual and abstain making the resulting classification of user queries more fine-grained.

Moreover, we introduce ORCAS-I, a new user intent classification dataset which extends the popular ORCAS dataset. The dataset is annotated using weak supervision with Snorkel. This approach enables obtaining labels for all 18M query-URL pairs. It can be a suitable resource for altering retrieval system results to match the type of information request from the user. For example, for transactional queries search engines can put heavier weight on results with commercial content or sponsored links. By contrast, providing commercial results for factual queries should be avoided. The general domain and the size of the dataset, together with the  taxonomy, allows multiple researchers to successfully train machine learning models to predict user intent to filter retrieval results. 

Besides the annotated dataset, we also release our labelling functions that can be highly useful for future application. This also makes it easy to improve the labelling functions using feedback from other researchers and release an updated version of labelled datasets.

In addition to the the weakly supervised dataset, we also publish a manually annotated subset that can be used for benchmarking the quality of machine learning models. We test the accuracy of our weakly supervised annotations and compare them to five benchmark models showing that none of them is able to significantly outperform Snorkel's output.

One limitation of our study is that we fine-tuned the labelling functions based on the ORCAS dataset characteristics, such as the most commonly visited domains. URLs from the United States are over-represented in the ORCAS dataset. Users searching with the same query in another country, would be re-directed to a different website based on their location (especially for queries regarding medical and legal advice). Because there were no other click log datasets to our availability, we were not able to generalise the labelling functions to location-dependent URLs.

Future work will focus on improving the labelling functions to reach better generalisation on datasets with other location-aware features, and also extending the taxonomy to cover the exploratory intent.


\begin{acks}
This work was supported by the EU Horizon 2020 ITN/ETN on Domain Specific Systems for Information Extraction and Retrieval -- DoSSIER (H2020-EU.1.3.1., ID: 860721).
\end{acks}

\bibliographystyle{ACM-Reference-Format}
\balance 
\bibliography{references.bib}


\end{document}